\documentclass[a4paper]{IEEEtran}

\usepackage{footnote}
\usepackage{nopageno}
\usepackage{float}
\usepackage{url}
\usepackage{amssymb,amsmath,amsfonts}
\usepackage{graphicx}
\usepackage{mathptmx}  
\usepackage{textcomp}
\usepackage{xcolor}
\usepackage[absolute,showboxes]{textpos}
\usepackage{multirow}
\usepackage{multicol}
\usepackage{array} 
\usepackage{cite}
\usepackage{algorithm,algorithmic}
\usepackage{amsmath}
\usepackage[cmintegrals]{newtxmath}
\usepackage[utf8]{inputenc}
\usepackage[T5]{fontenc}
\usepackage{tikz}
\usetikzlibrary{positioning}
\usepackage{color,soul}

\begin{document}
\pagenumbering{gobble}
\title{Destination-aware Adaptive Traffic Flow Rule Aggregation in Software-Defined Networks}
\author{
	\IEEEauthorblockN{Trung~V.~Phan, Mehrdad~Hajizadeh, Nguyễn Tuấn Khải, Thomas~Bauschert\IEEEauthorrefmark{1}\\}
     \IEEEauthorblockA{Technische Universit{\"a}t Chemnitz, Chair of Communication Networks, 09126 Chemnitz, Germany\\}
	\IEEEauthorblockA {Email: trung.phan-van | mehrdad.hajizadeh | tuan-khai.nguyen | thomas.bauschert@etit.tu-chemnitz.de}
    \thanks{\IEEEauthorrefmark{1}Corresponding author}
}

\markboth{Published in NetSys conference 2019}%
{Trung~V.~Phan \MakeLowercase{\textit{et al.}}: Destination-aware Adaptive Traffic Flow Rule Aggregation in Software-Defined Networks}

\maketitle

\begin{abstract}
In this paper, we propose a destination-aware adaptive traffic flow rule aggregation (DATA) mechanism for facilitating traffic flow monitoring in SDN-based networks. This method adapts the number of flow table entries in SDN switches according to the level of detail of traffic flow information that other mechanisms (e.g. for traffic engineering, traffic monitoring, intrusion detection) require. It also prevents performance degradation of the SDN switches by keeping the number of flow table entries well below a critical level. This level is not preset as a hard threshold but learned during operation by using a machine-learning based algorithm. The DATA method is implemented within a RESTful application (DATA App) which monitors and analyzes the ongoing network traffic and provides instructions to the SDN controller to adapt the traffic flow matching strategies accordingly. A thorough performance evaluation of DATA is conducted in an SDN emulation environment. The results show that---compared to the default behavior of common SDN controllers---the proposed DATA approach yields significant SDN switch performance improvements while still providing detailed traffic flow information on demand. 
\end{abstract}

\begin{IEEEkeywords}
Adaptive Traffic Aggregation, Resource Consumption, Network Statistics, Software Defined Networks.
\end{IEEEkeywords}

\section{Introduction}
Software Defined Networking (SDN) is a new networking paradigm which brings numerous advantages w.r.t. dynamic traffic control and management. The SDN concept overcomes the restrictions of legacy network architectures by decoupling the control and data plane, and handling the control plane in a centralized entity called SDN controller. The global network view of the SDN controller allows to enable a policy-based traffic management and a faster and more dynamic response to network state and traffic variations \cite{SDN}.

Several features \cite{RelatedWork} are already available in SDN networks such as mechanisms for traffic analysis, traffic flow management and resilience. Nevertheless, some challenges still remain to be addressed \cite{SDNChallenge}. In particular, adapting the granularity of traffic forwarding is an important issue. Most traffic management approaches rely on the default flow matching strategies of the available SDN controllers and therefore do not allow traffic flow handling with variable granularity. For instance, the \textit{Open Network Operating System} (ONOS) SDN controller \cite{onos}, by default, applies \textit{Reactive Forwarding} and uses the \textit{destination MAC address} for packet matching only. Hence, an incoming packet is matched to a flow entry by just using its layer 2 destination address. In order to change the flow matching scheme, an administrator has to manually set the respective $true$ or $false$ variables for the packet matching fields in the ONOS source code.

In this paper, we propose a destination-aware adaptive traffic flow matching (DATA) mechanism for SDN-based networks in order to adapt the number of flow table entries in SDN switches according to the level of detail of traffic flow information that other mechanisms (e.g. for traffic engineering, traffic monitoring) require. A RESTful application (DATA App) monitors and analyzes the network traffic and advises the SDN controller to adapt the matching strategy.
The paper is structured as follows. Section \ref{ProblemStatement} gives a short overview about common flow matching strategies in SDN-based networks. Related work is outlined in Section \ref{RelatedWork}. Section \ref{DATAProposal} explains the DATA solution in detail. The results of the performance evaluation are outlined in Section \ref{PerformanceEvaluation} and Section \ref{Conclusion} provides a short summary of our work.

\section{Common SDN Flow Matching Strategies and Their Implications}\label{ProblemStatement}

ONOS \cite{onos} provides high scalability and availability through its distributed architecture. It supports \textit{Reactive Forwarding} (fwd) and \textit{intent-based Reactive Forwarding} (ifwd) \cite{onos}. By default, the ONOS SDN controller provides flow rules using destination MAC addresses, i.e. only the destination MAC address is examined during the packet matching process and the other packet header fields like src/dst IP addresses or src/dst ports are not considered. We denote this flow matching strategy as MAC Matching Only Scheme (MMOS). However, by modifying the ONOS Reactive Forwarding configuration file, it is possible to add both the src/dst IP addresses and the src/dst ports to the matching fields - this flow matching strategy we denote as Full Matching Scheme (FMS). Similarly, OpenDaylight (ODL) \cite{odl} provides a default L2Switch service which applies MMOS. ODL also allows for another flow matching scheme which uses only destination IP addresses.

In MMOS the destination MAC address but no IP addresses and port numbers are included in the matching fields. Consequently, flow-table space is saved and the SDN switch might have a higher data plane forwarding performance (as the packet matching operation is quite fast). Another significant advantage of MMOS is that it reduces the workload of the SDN controller as less \textit{packet\_in} messages are generated. For example, MMOS does not care about new TCP requests to the same destination because it only checks the layer-2 addresses. However, the simple MMOS scheme may raise problems for other applications which want to monitor traffic flows in the network. For instance, an intrusion detection application requires detailed flow information to detect malicious traffic. As a result, an intrusion prevention application cannot issue the right policies for specific flows to prevent or mitigate unwanted traffic (e.g., DDoS traffic). In general, by using only MMOS, the ability to track and monitor network traffic for security or forensic analysis is limited.

In FMS, as the MAC address, IP address, and port number are used for packet matching, it is possible to classify traffic based on any individual field or combination of these fields. This fine-grained flow handling enables security or traffic engineering applications to have a closer view on the current network traffic. On the contrary, the number of \textit{packet\_in} messages to the SDN controller as well as the number of flow entries in a SDN switch is much higher than in the case of MMOS. This can result in significant degradation of the forwarding performance or even to a switch outage in case the maximum number of flow table entries is reached.

\section{Related Work}\label{RelatedWork}
Issues related to flow rule installation and management in SDN switches attracted a high interest in the SDN research community. A wide range of approaches to control TCAM (Ternary Content Addressable Memory) utilization were proposed with the primary target of flow rule compression or aggregation \cite{Stephens:2012,Leng:2015,Luo:2014,Mimidis:2016,Minnie:2017}. For example, the authors in \cite{Stephens:2012} argue that for simple packet forwarding rules based e.g. on MAC addresses or VLAN IDs only cheap SRAM memory is sufficient. Only more complex matching rules (with more matching fields) might require the use of fast but expensive TCAM memory. Thus the amount of TCAM memory in SDN switches can be significantly reduced. The solutions outlined in \cite{Leng:2015} and \cite{Luo:2014} apply the concept of flow rule aggregation by restructuring the matching fields. By that the number of flow rules can be significantly reduced. Another approach for dynamic flow matching was proposed in \cite{Mimidis:2016} where the matching policy includes the DSCP (Differentiated Services Code Point) values for different traffic types. Rifai et al. introduced the MINNIE framework \cite{Minnie:2017} for flow table compression using wildcard rules. The mentioned approaches only focus on flow table size reduction and on increasing the data plane forwarding performance but---contrary to our approach---do not consider the possibility of adaptively changing the granularity of flow matching (and thus the flow table size) depending on the level of detail of traffic flow information that other mechanisms (traffic engineering, traffic monitoring) demand. 

\section{Destination-aware Adaptive Traffic Flow Rule Aggregation Mechanism} \label{DATAProposal}
\subsection{DATA Architecture}\label{SoftwareDesignArchitecture}
\begin{figure}
\centering
\includegraphics[width=0.49\textwidth]{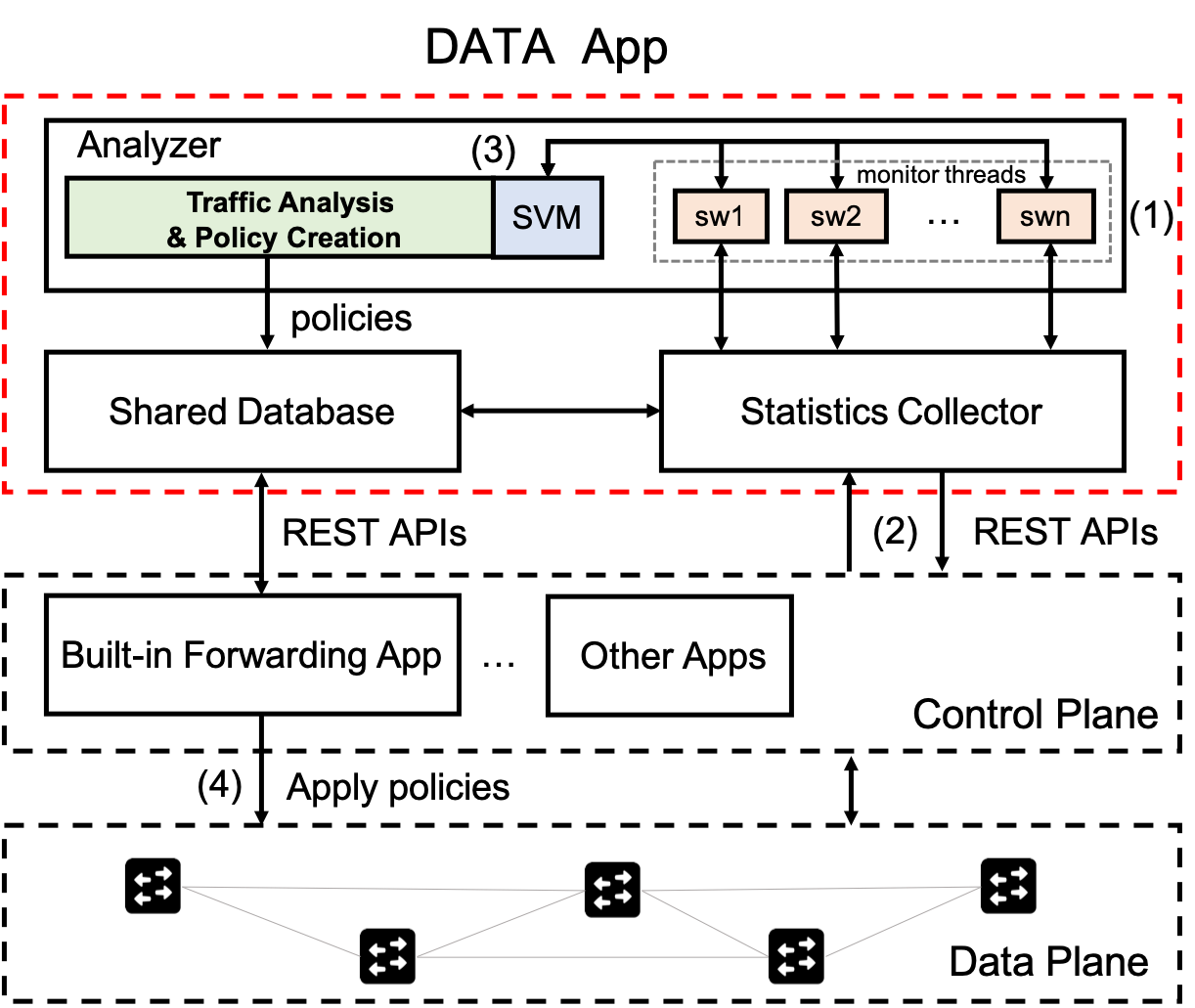}
\caption{DATA Architecture}
\label{fig:DATA}
\end{figure}

Fig. \ref{fig:DATA} provides an overview of the extended SDN control plane. It comprises the default SDN control plane with a built-in forwarding application and our novel DATA App. Detailed information about the relevant components is provided below. 

\subsubsection{Built-in Forwarding Application}
Most of the common SDN controllers \cite{onos,odl} provide a built-in forwarding application with basic functionality to allow the creation of flow rules which are then downloaded to the SDN switches. In this work, we propose to add REST API interfaces to the built-in forwarding application to have a secure communication channel to the DATA App. The channel is used to share and exchange network information and control instructions with the DATA App via the shared database. Initially the DATA App instructs the build-in forwarding application to apply the FMS strategy.

\subsubsection{DATA App}
The DATA App consists of the following three main components: The Statistics Collector periodically gets information about the traffic flows (e.g. MAC/IP addresses, packet counts) traversing the SDN switches from the SDN controller via a RESTful API \cite{onos,odl}. It sends the collected statistical information to the Analyzer and stores it in the shared database. The \textit{Shared Database} is accessible from three agents: Built-in forwarding application, Statistics Collector and Analyzer. The \textit{Analyzer} controls the change of the flow matching schemes in the SDN network. It receives flow statistics information from \textit{monitor threads} of SDN switches via the Statistics Collector and identifies the SDN switches which are subject to performance degradation and finds out the \textit{destination} hosts (via applying Algorithm \ref{FMStoMMOS}, see below) whose associated flows are most critical to the switch performance. In order to anticipate the switch performance degradation well before it occurs and to trigger the flow matching scheme change in time we apply a 2-dimensional Support Vector Machine (SVM) \cite{SVMIntro} learning algorithm which is well known in the machine learning research community for its very good practical results \cite{AIGoogle}. After checking the switch performance, the Analyzer co-operates with the built-in forwarding application of the SDN controller to conduct actions regarding the flow matching schemes of all flows related to the previously identified destination hosts. 
 
\begin{figure}
\centering
\includegraphics[width=0.49\textwidth]{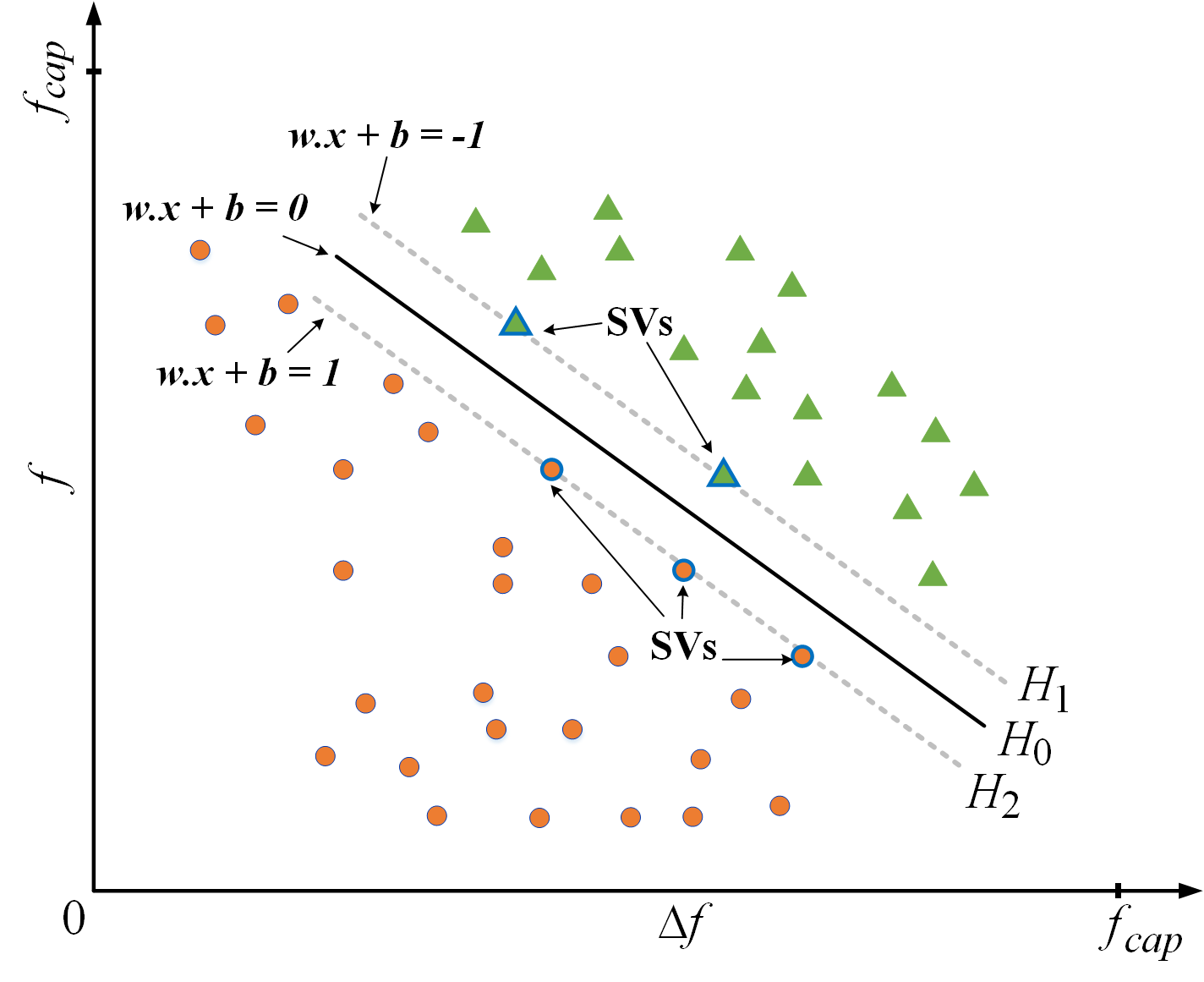}
\caption{Support Vector Machine principle}
\label{fig:SVM}
\end{figure}

The SVM algorithm in the DATA App works as follows. In general we have a linearly separable data set for training  $D = \left\{\left({x}_1,\textit{y}_1\right), \left({x}_2,\textit{y}_2\right),\dots,\left({x}_N,\textit{y}_N\right)\right\}$, where ${x}_i \in \mathbb{R}^{n}$ and $y_{i} \in \left\{+1,-1\right\}$. In this work ${x}_i$ represents the tuple ($f$,$\Delta f$). Here $f$ ($0 \leq f \leq f_{cap}$) is the current total number of flow entries and $\Delta f$ denotes the change of the total number of flow entries in a SDN switch between two consecutive observations. The value of $y_i$ represents the status of the switch in the $i$th observation period: $y_i$ = +1 denotes a good switch state while $y_i$ = -1 indicates a performance degradation (due to e.g. errors or exceptions). The SVM algorithm operates in two main phases: training and mapping. In the training phase, the SVM algorithm takes data samples from the data set $D$ and tries to find three hyperplanes $H_0$, $H_1$ and $H_2$ as follows:
\begin{equation}
    \left\{ 
    \begin{array}{ll}
    H_0 = \left\{{x} \in \mathbb{R}^{n} : \left\langle{w},{x}\right\rangle + \textit{b} = \text{0},\; {w} \in \mathbb{R}^{n},\;\textit{b} \in \mathbb{R}\right\} & 
    \\ H_1 = \left\{{x} \in \mathbb{R}^{n} : \left\langle{w},{x}\right\rangle + \textit{b} = \text{-1},\; {w} \in \mathbb{R}^{n},\;\textit{b} \in \mathbb{R}\right\}  & 
    \\ H_2 = \left\{{x} \in \mathbb{R}^{n} : \left\langle{w},{x}\right\rangle + \textit{b} = \text{+1},\; {w} \in \mathbb{R}^{n},\;\textit{b} \in \mathbb{R}\right\} &
	\end{array}\right.
\end{equation}
see Fig. \ref{fig:SVM}. The region bounded by the $H_1$ and $H_2$ hyperplanes is called the margin in which no data samples of the training set are allowed to be in. The $H_0$ hyperplane lies in the middle between $H_1$ and $H_2$. Note, that the chance of finding more than three hyperplanes to separate two data groups is relatively high. However, there is only one optimal solution that maximizes the margin between $H_1$ and $H_2$. The task is to find the values $w$ and $b$ so that the margin is maximum. The solution of this optimization problem is described in \cite{SVMIntro}. The distance of $x_i$ to the hyperplane $H_0$ is defined as follows: 
\begin{equation}
    \frac{y_i (\left\langle{w},{x}\right\rangle + \textit{b})}{\parallel w \parallel} \geq 1,
\end{equation}
 where the sign of $y_i$ indicates the data group to which $x_i$ belongs. In the mapping phase, for a new sample $x_i$ it is checked to which of the two data groups separated by the hyperplane $H_0$ it belongs. This is done according to the result of the function $F(x)=sign(\left\langle{w},{x}\right\rangle + \textit{b})$. If $F(x)$= +1 the sample $x_i$ is assigned to the data group representing a good switch performance status, otherwise ($F(x)$= -1) the sample $x_i$ is assigned to the data group representing a performance degradation of the switch. Interested readers are referred to \cite{SVMIntro,OpenFlowSIA,SVMSOM,SVMSOM-IEEEAccess} for a detailed explanation of the SVM algorithm. 

The reasons for choosing the tuple ($f$,$\Delta f$) for evaluating the forwarding performance of a SDN switch are as follows: The effort for flow searching and matching within a SDN switch is proportional to the number and matching fields of flow entries. Moreover, a SDN switch has a maximum capacity ($f_{cap}$) for storing the flow entries. Accordingly, the change of the number of flow entries indicates the control plane load (w.r.t. $of\_mod$ and $of\_removed$ messages sent between SDN controller and SDN switch) affecting both the SDN switch and the SDN controller.

\begin{figure}
\centering
\includegraphics[width=0.48\textwidth]{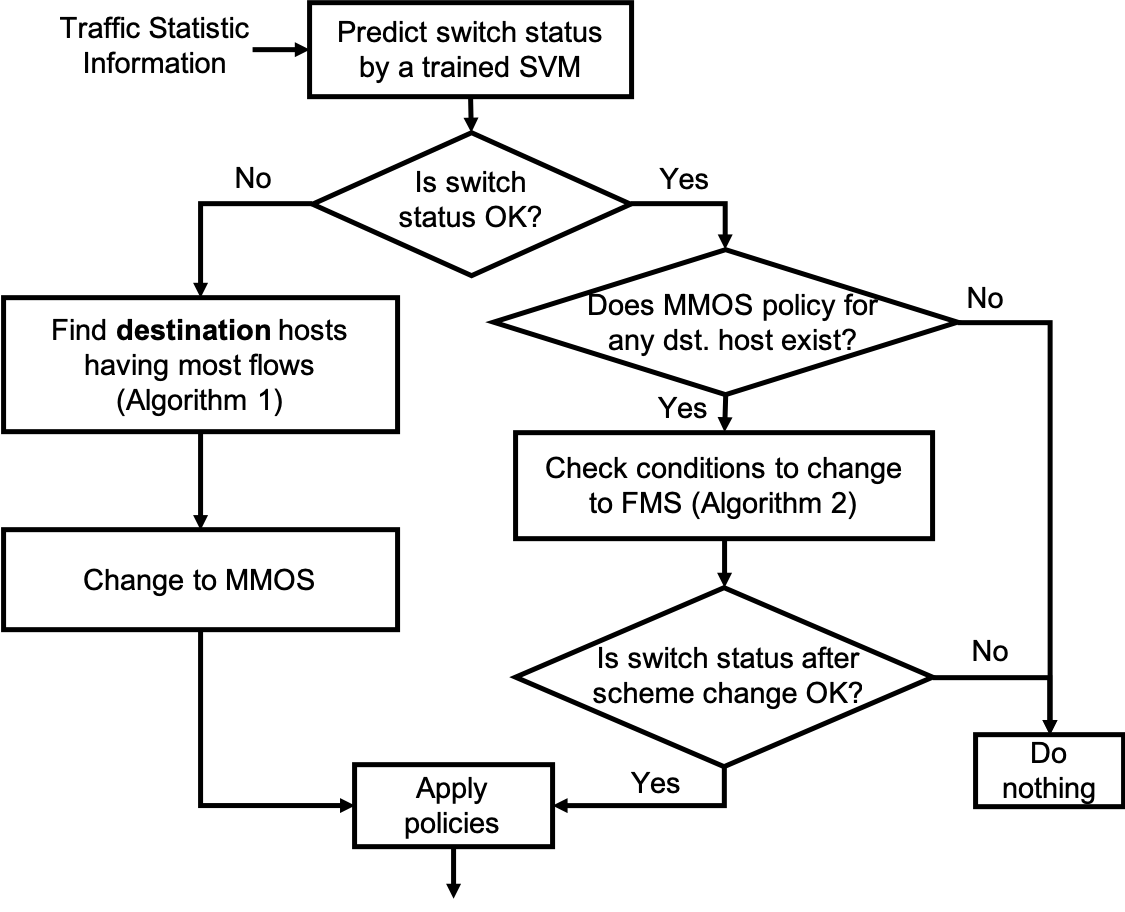}
\caption{Traffic analysis and policy creation at the Analyzer}
\label{fig:DATAWorkflow}
\end{figure}

\begin{algorithm}
\caption{Identification of the destination hosts whose flows are most critical to the performance of switch $i$}
\label{FMStoMMOS}
\begin{algorithmic}
\STATE \textbf{Input}: $S_i$ = $\left\{(h_{1}, f_{1}), (h_{2}, f_{2}), ..., (h_{k}, f_{k})\right\}$: set of destination hosts and respective number of flow entries associated with these hosts in switch $i$, $f_i$ = $\sum_{c=1}^{k} f_{c}$: total number of current flow entries in switch $i$, $p$ = 1: first index
\STATE \textbf{Output}: $H_{j}$ = $\left\{ \right\}$: set of destination hosts
\STATE \textbf{begin}
\STATE Sort $S_i$ in descending order of the current flow $f_c$ (from highest to lowest numbers)
\LOOP
\STATE $H_j.append(S_{i}$[$p$])
\STATE $f_{remaining} = 1 + \sum_{c=p+1}^{k} f_{c}$ \COMMENT{One MMOS flow entry is installed in switch $i$}
\STATE $\Delta f = f_i-f_{remaining}$ \COMMENT{Delete $\Delta f$ flow entries in switch $i$}
\STATE ${x}$ = ($f_{remaining}$, $\Delta f$)
\STATE $sign = \textit{SVM}\left({x}\right)$ \COMMENT{Feed $x$ into SVM}
	\IF {$sign$ = +1}
		\STATE \textbf{break}
        \COMMENT{Switch $i$ can handle $f_{remaining}$ entries}
	\ELSE
    	\STATE $p$ = $p$+1
    	\COMMENT{Switch $i$ cannot handle $f_{remaining}$ entries}
	\ENDIF
\ENDLOOP
\STATE \textbf{return} $H_j$
\end{algorithmic} 
\end{algorithm}

\subsection{Operational Workflow}
Initially, the Statistics Collector sends a request to the SDN controller to ask for network topology information. Then, it launches a monitor thread for each connected SDN switch (step (1) in Fig. \ref{fig:DATA}). The monitoring information is stored in the shared database. Meanwhile, the Analyzer activates the SVM engine and performs the training phase using a pre-prepared training data set, see Section \ref{scenariosetup}. Next, the Statistics Collector gets traffic flow statistics from the connected SDN switches (step (2) in Fig. \ref{fig:DATA}). In regular time intervals (observation period) - for each SDN switch - a monitor thread counts the total number of current flows and measures the flow number changes in order to provide the tuple ($f$,$\Delta f$) to the SVM engine within the Analyzer. The Analyzer then conducts traffic analysis and policy creation for each SDN switch (step (3) in Fig. \ref{fig:DATA}). 

As illustrated in Fig. \ref{fig:DATAWorkflow}, in case the SVM engine detects a performance degradation of a switch $i$, the destination hosts whose associated flows are most critical to the performance of switch $i$ (i.e. have the most flow entries in switch $i$) are identified by applying Algorithm \ref{FMStoMMOS}. Subsequently, the Analyzer instructs the built-in forwarding application to send \textit{of\_mod} messages to remove all full-matching flow entries in the flow-table of switch $i$ and replace them by MAC matching only flow entries, i.e. to perform a change from FMS to MMOS (step (4) in Fig. \ref{fig:DATA}). Furthermore, the DATA App monitors the number of incoming packets per second ($R_{pkt}$) in switch $i$ individually for all flows, for which the matching scheme change to MMOS is applied.

In case no performance degradation is detected for switch $i$ it is checked whether there exists a MMOS policy for any destination hosts. If an MMOS policy is found, then Algorithm \ref{MMOStoFMS} is applied to check the conditions for a change to the FMS strategy (step (4) in Fig. \ref{fig:DATA}).

\begin{algorithm}
\caption{Identification of the MMOS flows/destination hosts related to switch $i$ for which changing back to FMS is feasible}
\label{MMOStoFMS}
\begin{algorithmic}
\STATE \textbf{Input}: $S_i = \left\{(h_{1}, R_{pkt_1}), (h_{2}, R_{pkt_2}), ..., (h_{m}, R_{pkt_m})\right\}$: set of destination hosts and respective packet rate of MMOS flows associated with these hosts in switch $i$, ($f_i$, $f_{cap}$): total number of current flow entries and maximum number of flow entries in switch $i$, $f_{extra}$: number of flow entries that might be added in switch $i$
\STATE \textbf{Output}: $H_{j} = \left\{ \right\}$: set of destination hosts
\STATE \textbf{begin}
\FOR {$s = 1$; $s \leq m$; $s$++}
    \STATE $f_{extra}$ = \textit{idle\_timeout}*$R_{pkt_{s}}$ \COMMENT{Worst case assumption: each packet is associated with a new entry in switch $i$}
  \IF {($f_{extra} + f_i) < f_{cap}$}
    \STATE $H_{j}$.append[$h_s$]
  \ELSE
    \STATE \textbf{continue}
  \ENDIF
\ENDFOR
\STATE \textbf{return} $H_{j}$

\COMMENT{\textit{idle\_timeout} is a period of time set in a flow entry. If there is no more incoming packets that matches to the flow entry since last matched packet, then the flow will be removed after \textit{idle\_timeout} seconds.}
\end{algorithmic} 
\end{algorithm}

Algorithm \ref{MMOStoFMS} works as follows: for each switch (which has MMOS applied) the identified flows respectively destination hosts are sorted in ascending order w.r.t. the packet rate $R_{pkt}$ of the identified flows. For the \textit{first} destination host in the list a flow matching scheme change to FMS is applied in the switch. For that, the built-in forwarding application is instructed to send \textit{of\_mod} messages to remove all MMOS flow entries in the affected switches (step (4) in Fig. \ref{fig:DATA}). The decision about changing the flow matching strategy happens once per observation period (which is set to 3 seconds in our implementation). By that strategy we increase the number of flow entries in a switch only moderately (per observation period) and avoid large variations in the number of flow entries.

\begin{figure}
\centering
\includegraphics[width=0.5\textwidth]{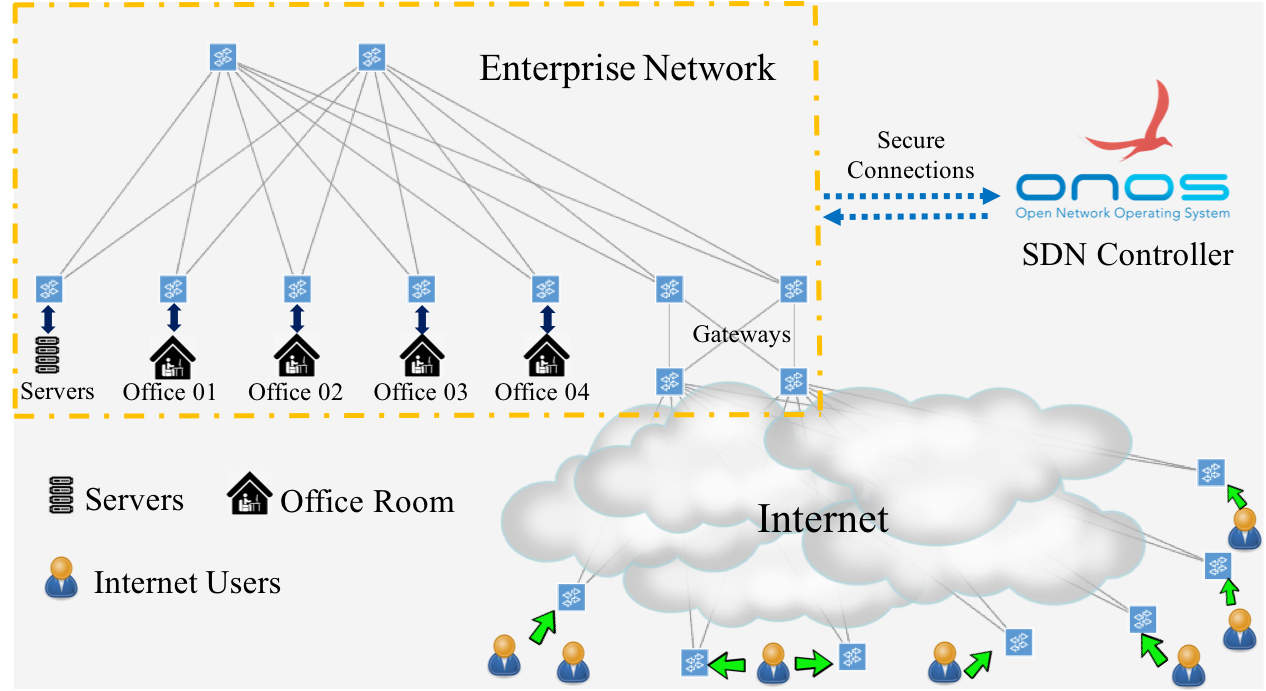}
\caption{DATA deployment example (Enterprise SDN network emulated with MaxiNet)}
\label{fig:DATADeployment}
\end{figure}

\section{Performance Evaluation}\label{PerformanceEvaluation}

\begin{figure*}
\begin{multicols}{3}
    \includegraphics[width=\linewidth]{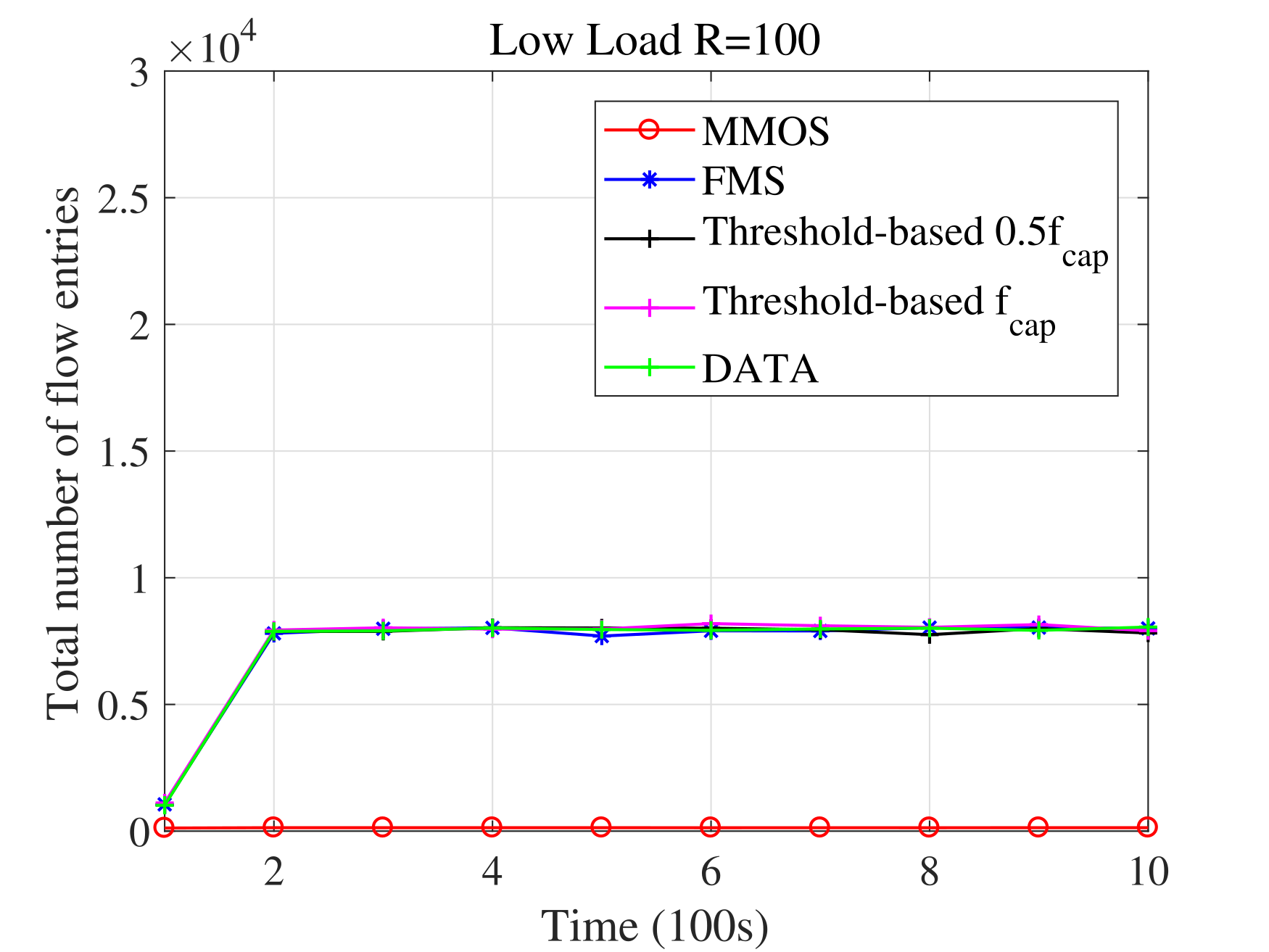}\par
    \includegraphics[width=\linewidth]{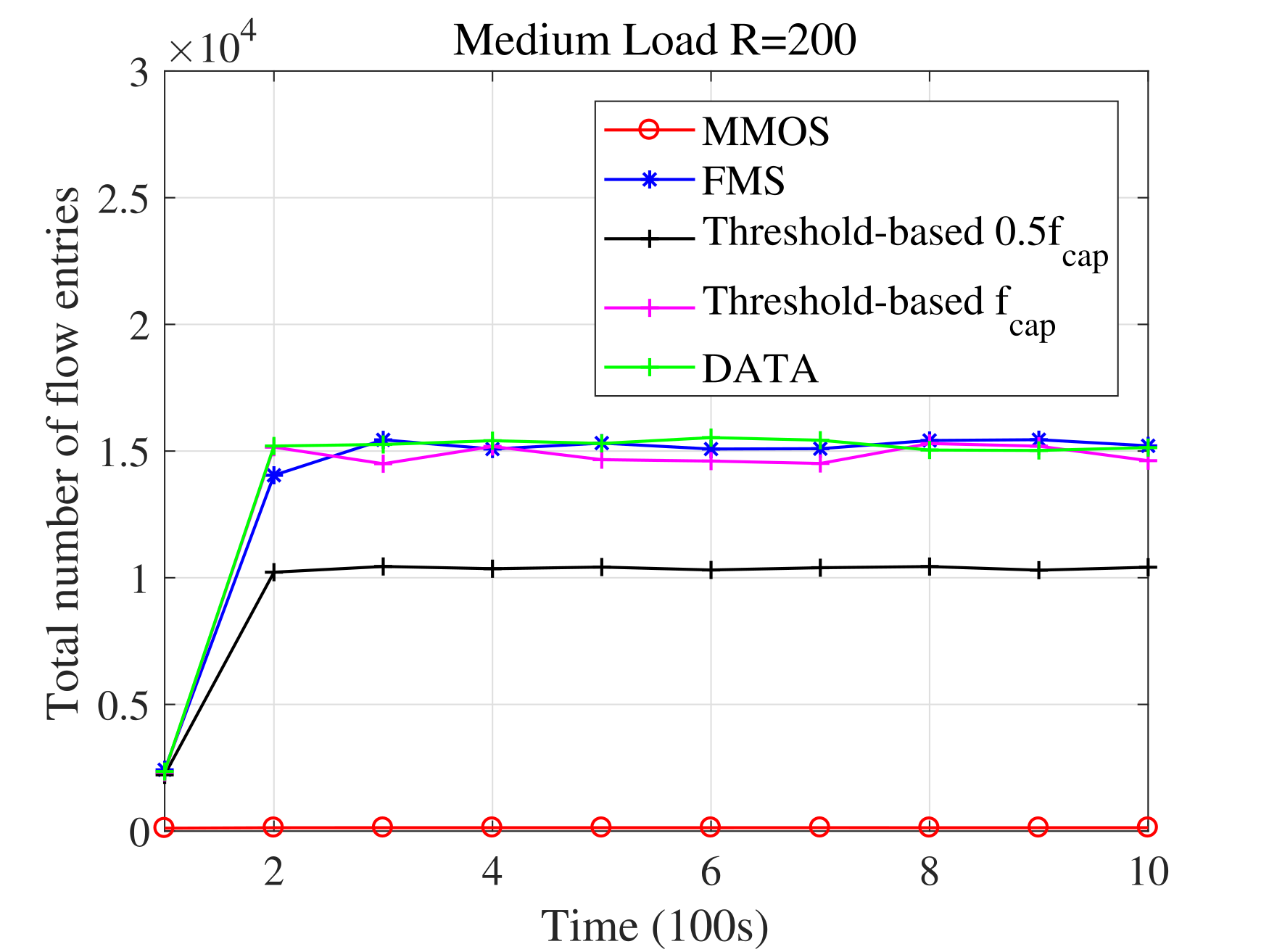}\par
    \includegraphics[width=\linewidth]{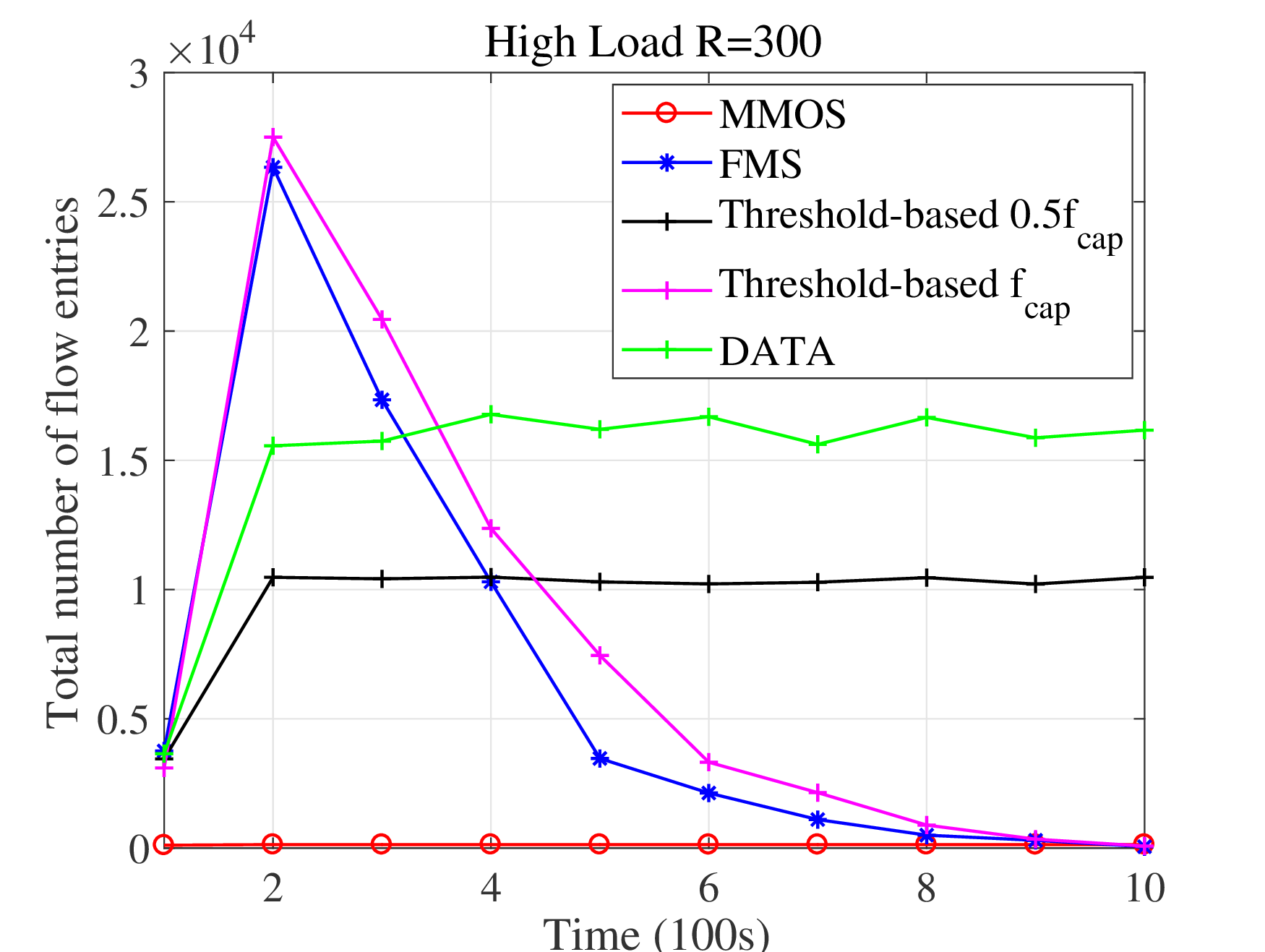}\par
\end{multicols}
\caption{Total number of flow entries in the enterprise SDN network over time for three different traffic loads $R$}
\label{fig:FlowNumber}
\end{figure*}

\subsection{Scenario Setup}\label{scenariosetup}
The MaxiNet framework \cite{MaxiNet} is applied to emulate an enterprise SDN network comprising several SDN switches (realized via OpenvSwitch), 96 enterprise hosts (24 hosts per office), 24 enterprise servers (within one server rack) and a connection to the Internet (see Fig. \ref{fig:DATADeployment}). The emulated enterprise SDN network runs within one Linux PC and is controlled by a remote ONOS SDN controller running on another PC. The Internet hosts are emulated on a third Linux PC. Enterprise and Internet hosts are running within Linux containers using Ubuntu images, and the servers are running within Linux containers using Apache Web server images. For a convenient configuration we place both the ONOS SDN controller and the DATA App on the same Linux PC.

Initially, for training the SVM we generate traffic from enterprise and Internet users towards the enterprise servers and among enterprise users and apply the FMS scheme for these traffic flows. Contrary, for the traffic flows from the servers towards the enterprise and Internet users (response traffic) we apply MMOS. We monitor any errors or exceptions indicating that switches cannot handle new flow requests or that switches are disconnected from the SDN controller. We capture the tuple ($f$,$\Delta f$) at all SDN switches and set \textit{sign} = -1 if the switch performance degrades (due to errors or exceptions); otherwise, we set \textit{sign} = +1. These tuple samples are then used for training the SVM in the Analyzer. We observe that an SDN switch starts getting overloaded or cannot handle new flow rules when the current number of flows is around 3000 ($f_{cap}$). Setting the \textit{idle\_timeout} value (after which the flow entries are removed) to 10 seconds, the safety threshold for the packet rate $R$ a switch can handle is 300 packets per second assuming that each packet belongs to a different flow (worst case assumption). Accordingly, for the traffic generation we divide R into three levels: low ($R$=100), medium ($R$=200) and high ($R$=300).

In our performance analysis we carry out several experiments with different flow matching strategies: MMOS only, FMS only, Threshold-based (considering the number of flow entries, $f_{thres}$, as threshold in the DATA App without applying the SVM engine) and adaptive (DATA App with SVM engine). The ONOS controller applies \textit{Reactive Forwarding}. Furthermore, we implement an IDS application to detect abnormal traffic. The IDS application is based on a Self Organizing Map algorithm \cite{SOM} which classifies traffic by the 4-tuple \textit{average number of packets per flow}, \textit{average number of bytes per flow}, \textit{average duration per flow} and \textit{percentage of pair-flows}. For the performance analysis we generate traffic from enterprise and Internet users towards the enterprise servers and among enterprise users with three different load levels $R$ = (100, 200, 300). During the experiments we trace the total number of flow entries in the enterprise SDN network and extract the average number of \textit{packet\_in} messages per second to the ONOS controller. 

\subsection{Result Analysis}
\subsubsection{Total number of flow entries in the enterprise SDN network}
 Fig. \ref{fig:FlowNumber} shows that the MMOS scheme naturally accounts for a very small amount of flow entries in all cases, and that the FMS, the DATA and Threshold-based strategies are not much different for low and medium traffic load. In case of high traffic load, the total number of flow entries is still beyond a critical level both for the DATA as well as the Threshold-based scheme (with $f_{thres}=0.5f_{cap}$). This is due to the fact that despite increasing traffic flows, these two mechanisms significantly reduce the number of new flow entries by changing to MMOS in order to prevent performance degradation of the SDN switches. Contrary the FMS and the Threshold-based scheme (with $f_{thres}=f_{cap}$) continue to generate more and more flow entries which quite soon has negative effects (errors and exceptions) on both the built-in forwarding application of the SDN controller and some SDN switches leading to a gradual performance degradation. Finally the SDN controller and some switches suspend their operation. A low total number of flow entries remains due to the few switches which are still operational.

\begin{table}
\caption{Average \textit{packet\_in} rate (pkts/s) to the SDN controller for different traffic flow rule aggregation schemes and traffic loads $R$}
\label{AveragePacketin}
\centering
\begin{tabular}{|c|c|c|c|c|c|}
\hline
\bfseries &&& Threshold-& Threshold-&\\
Schemes&MMOS&FMS&based&based&DATA\\
&&&(0.5$f_{cap})$& ($f_{cap}$)&\\
\hline
$R$=100&0.066&187.33&180.33&183.33&184.33\\
\hline
$R$=200&0.066&369.33&340.33&368.33&363.33\\
\hline
$R$=300&0.066&562.33&387.33&553.33&393.33\\
\hline
\end{tabular}
\end{table}

\subsubsection{Average \textit{packet\_in} message rate to the ONOS controller}
In Table \ref{AveragePacketin} the average number of packets per second (\textit{packet\_in} rate) arriving at the built-in forwarding application quering for new flow rules is shown. It can be seen that, contrary to the FMS and the Threshold-based scheme (with $f_{thres}=f_{cap}$), for all traffic load levels, the Threshold-based scheme (with $f_{thres}=0.5f_{cap}$) and the DATA scheme allow the ONOS controller to have an acceptable \textit{packet\_in} rate and guarantee that the SDN switches are not getting degraded. Besides, they significantly reduce the workload of the built-in application due to the lower number of new flow queries.

\subsubsection{Errors and exceptions}
An important criterion for the performance evaluation of the DATA approach is the time until an error or exception (observed by the ONOS) occurs because of an overloaded SDN switch in the network. Our results show that the FMS and the Threshold-based scheme (with $f_{thres}=f_{cap}$) cause \textit{disconnected} channels errors and \textit{FlowRuleManager} exceptions in the ONOS controller after 7 to 10 seconds since the high traffic load ($R$ = 300) is generated. For the other traffic load cases no errors and exceptions are observed.

\begin{table}
\caption{Detection rate (\%) of our IDS application for different traffic flow rule aggregation schemes and traffic loads $R$}
\label{IDSDetectionRate}
\centering
\begin{tabular}{|c|c|c|c|c|c|}
\hline
\bfseries &&& Threshold-& Threshold-&\\
Schemes&MMOS&FMS&based&based&DATA\\
&&&(0.5$f_{cap})$& ($f_{cap}$)&\\
\hline
$R$=100 & 0.0 & 98.6 & 97.8 & 97.5 & 98.8 \\
\hline
$R$=200 & 0.0 & 98.7 & 82.45 & 97.9 & 97.5 \\
\hline
$R$=300 & 0.0 & 0.0 & 81.23 & 0.0 & 97.8\\
\hline
\end{tabular}
\end{table}

\subsubsection{Detection rate of the IDS application}
We assume that attackers from both the enterprise and the Internet launch a DDoS TCP SYN flooding attack to the enterprise Web servers. We measure the DDoS attack detection rate of our IDS application. The results in Table \ref{IDSDetectionRate} show that there is no alert in case of MMOS for all traffic loads because all traffic towards the Web servers is grouped into one flow entry at all switches. Hence, the IDS application can not recognize the DDoS attack. In case of low and medium load, with FMS, Threshold-based (with $f_{thres}=f_{cap}$) and DATA a TCP SYN flooding attack is detected by the Self Organizing Map algorithm (with quite similar detection rate). The Threshold-based scheme (with $f_{thres}=0.5f_{cap}$) accounts for a less attack detection rate due to the fact that traffic flows towards the Web servers are handled with MMOS whenever the threshold is reached. In our DDoS attack scenario the attacker tries to send as fast as possible TCP segments with different source TCP ports to the victim (Web servers). This leads to the installation of new flows entries in the SDN switches. Therefore, it is easy for the IDS application to gather traffic flow information and detect the attack. In the high load case and for FMS as well as for the Threshold-based (with $f_{thres}=f_{cap}$) scheme, the operations of the SDN controller and some switches are suspended. That makes the IDS application unable to gather traffic information from the SDN controller and detect the attack. On the contrary, with DATA, all SDN network components stay operational and the IDS application can gather detailed information about the new traffic flows and thus detect the TCP SYN flooding attack. DATA yields a detection rate that is 16.5\% higher compared to the Threshold-based scheme (with $f_{thres}=0.5f_{cap}$). Consequently, our novel DATA solution can efficiently provide useful information for security analysis avoiding the drawbacks of the other flow matching schemes.

\section{Conclusion}\label{Conclusion}
In this paper, we propose a destination-aware adaptive traffic flow rule aggregation mechanism (DATA) to adapt the number of flow table entries in SDN switches according to the level of detail of traffic flow information that other mechanisms (e.g. for traffic engineering, traffic monitoring, intrusion detection) require and at the same time prevent SDN switch performance degradation. Our performance evaluation proves that the DATA solution outperforms legacy flow rule matching schemes. In our future work, we are going to adapt DATA as integrated application for common SDN controllers.

\section*{Acknowledgment}
This work has been performed in the framework of the Celtic-Plus project SENDATE Secure-DCI, funded by the German BMBF (ID 16KIS0481).

\bibliographystyle{ieeetr}
\bibliography{resource.bib}
\end{document}